# Thermal noise induced stochastic resonance in self organizing Fe nanoparticle system

Satyendra Prakash Pal[1,2*] and P Sen[1]

[1]School of Physical Sciences, Jawaharlal Nehru University, New Delhi-110067, India

[2]Department of Physical Sciences, Indian Institute of Science Education and Research, Knowledge city, Sector 81, SAS Nagar, Manauli-140306, Punjab, India

[*]email: sppal85@gmail.com

**Abstract**

The natural world is replete with examples of multistable systems, known to respond to periodic modulations and produce a signal, which exhibits resonance with noise amplitude. This is a concept not demonstrated in pure materials, which involve a measured physical property. In a thermoremanent magnetization experiment with a common magnetic material, Fe, in the nanoparticulate form, we establish how magnetization in a system of dilute spins during dissipation of stored magnetic energy, breaks up into spontaneous oscillatory behavior. Starting at 175 K and aided by temperature (stochastic noise) the oscillation amplitude goes through a maximum, reminiscent of stochastic resonance. Our observation of thermal noise induced coherent resonance is due to intrinsic self-organizing magnetic dynamics of the Fe nanoperticle system without applying any external periodic force. These results yield new possibilities in design of magnetic materials and a platform to understand stochastic interference and phase synchronization in neural activity, as models for neural communication.

## Introduction

Stochastic resonance is a natural phenomenon observed often in large systems whereby a signal is either produced or sensed, significantly better, in the presence of optimal noise. Prey capture by paddlefish (1-3), ecological time series (4), human brain (5), cell biology (6,7), neural models (8,9) and neural populations (10–12) are such examples of large systems where a measureable quantity, naturally representative of the system, when monitored in the presence of periodic forcing is seen to benefit from stochastic noise. Magnetic materials figure only sparsely in this list. For example, stochastic resonance has been demonstrated in Bi-substituted ferrite-garnet films (13) and also in yttrium-iron garnet spheres (14), where the magneto-optic signal was shown to go through resonance in the presence of an external modulation. Low dimensional magnetic materials are ideal bistable systems as the magnetic moment cannot maintain anisotropy and gets rattled by temperature. Indeed, magnetic stochastic resonance was predicted theoretically for the interdomain magnetization tunneling in uniaxial ferromagnets (15, 16), for single-domain uniaxial superparamagnetic particles (17, 18), and in an assembly of single-domain ferromagnetic particles dispersed in a low-concentration solid phase (19). There has been however, no demonstration of stochastic resonance in any such material, which includes materials demonstrating superparamagnetic behavior.

Two reasons can be identified as basis for the failure. To demonstrate stochastic resonance a system comprised of several microscopic elements, must have the following characteristics. (i) It should have strong local moments, such that small but finite deviations from the fixed point can produce large signals so that the overall system can be nonlinear and (ii) the system must be autonomous, i.e., there must be no external driving present, as external driving will force an averaging effect on the bistable system (consisting of 2 fixed points), from several randomly oriented anisotropy axes. These observations are particularly true for granular nanoparticulates. Spin relaxation experiments reported in the literature certify first of these statements. Bedanta and co-workers (20) have studied spin relaxation behaviour in a system of granular nanoparticles and observe that at low concentrations, the spins relax exponentially while at higher concentrations, spin-spin interactions give rise to power law decay in the magnetization, typical of a supercooled system demonstrating glassy behaviour. Thus existing systems are not sufficiently strong, not sufficiently nonlinear and interacts only at close proximity.

We prepare our superparamagnetic Fe nanoparticles in such manner that a single domain particle, measuring about 7.5 nm across, is further divided into smaller domains, without boundaries. This is normally not possible, as small particles, once formed, are instantaneously capped. However when a nonequilibrium process such as electro-explosion of wires (21) is employed to achieve nanoparticle formation, lattice disruptions occur in some materials with low electronic conductivity (or higher electron-phonon coupling, as electrons are unable to carry away sudden excess kinetic energy). Fe is one such material which exhibits this behavior (22, 23).

In this article, we have presented the thermal noise induced coherent resonance behaviour in the Fe nanoprticles. These observations are self generated from the nanopartilce system without any external periodic driving force.

**Experimental methods:**

Magnetic relaxation data at various temperatures, starting from 100K to 300K, were collected by using a commercial Quantum Design SQUID magnetometer (MPMS-XL) with a sensitivity of $10^{-8}$ emu. Magnetic relaxation curves were obtained at various temperatures after ramp up of the sample temperature to 330K followed by cool down to the pre-assigned temperatures, in the presence of an applied magnetic field (2T). The magnetic field was then reduced to zero in the 'no overshoot' mode in 300s, and then magnetic relaxation data was collected with 10s time delay between two data points. A "Reciprocating sample transport option" (RSO) was used in these measurements to achieve better sensitivity.

**Results and discussion:**

We have prepared our superparamagnetic Fe nanoparticles, measuring about 7.5 nm across, by using a nonequilibrium process of electro-explosion of wires (21). Due to higher value of electron-phonon coupling for Fe, as electrons are unable to carry away sudden excess kinetic energy, disrupted lattice structure for the nanoparticles has been achieved. The resulting nanoparticle lattice is schematically depicted in Fig. 1.

The regions with lattice distortions (shown as red dots with a cross to identify lattice distortion) delineate ordered regions (colored blue). This results in frustrated magnetism, where a disruption in long range lattice order is associated with disrupted magnetic order. Magnetic moments, supported by magnetic frustration, have been shown to be considerably stronger than the rest of a lattice (24). When a thermoremanent magnetization (TRM) measurement is carried out to track relaxation dynamics of the spins, they behave like noninteracting spins dominated by a relaxation term $\exp[-t/\tau]$. This is schematically depicted as well in Fig. 1a, whereby till $t=t_0$, the relaxation is purely exponential and beyond this it deviates from exponential behavior in the details as shown here (oscillatory, with an overall power law decay).

In a recent paper (25), we have correlated the occurrence of an intermixture of lattices to the amplitude of the oscillatory component. Using an anneal procedure designed for this purpose; we were successful to partly smooth out the disordered lattice and obtain a situation depicted in Fig. 1b. The disordered sites decrease in number, with a simultaneous impact (decrease) in the amplitude of the oscillating magnetic moment. In order to understand the complete character of these oscillations, we carry out a similar measurement at various temperatures, namely, 175K, 200K, 250K and 300K. In the SQUID magnetometer used for this purpose, data was collected with a time delay of 10s, so as to gather a large information data set. Care was taken to rule out possibilities of instrument error by comparing this data with the ones collected every 200s. Major frequency components are seen in the latter while minor ones are present in addition to the major ones, for the former.

To start with, we show in Fig. 2, the magnetic relaxation data collected at 100K and 175K. The data at 100K (Fig. 2a) represents exponential decay with a combination of 2 relaxation times, $\tau_{1(100K)}$= 744s and $\tau_{2(100K)}$= 7491s, which identify the spins as noninteracting. At 175K, the initial relaxation is exponential with a relaxation time $\tau_{1(175K)}$= 262s, however, beyond 4000s oscillations appear. The initial relaxation itself is faster as seen here with $\tau_{1(175K)} < \tau_{1(100K)}$. This releases the magnetic moments which were forced into orientation in a particular direction due to the 2T external field in the TRM measurement, to quickly achieve a state of chaotic dynamics and possibilities of noise induced order (26) if the other conditions as listed above, are satisfied.

As the oscillations are a result of spins released, now excitable, the observed amplitude can be conceived as a signal, proportional to the number of participating spins. In order to get some meaningful result out of the oscillatory part of the data in Fig. 2b, we perform a background subtraction procedure by first smoothing the data with a spline fit which draws an average line through the selected range. The deviations from this line are the perturbations seen by the SQUID magnetometer detectors, $\Delta M$, which would have otherwise followed a relaxation behavior as seen prior to t=4000s.

The resultant data retrieved this way by a suitable background subtraction at various temperatures, 175K, 200K, 250K and 300K are plotted in Fig. 3. A prominent evolution of periodic behavior is observed with increasing temperature, T, by visual inspection of the background subtracted data. The other observation is a clear increase in the signal amplitude, or extent of signal excursion, as one goes to 300K. These signals are autonomous or self-generated, and as seen here, aided by the temperature at which the experiment is carried out.

Next, in Fig. 4, we presented Fourier analysis (FFT) for the relaxation data taken at 175K, 200K, 250K and 300K. FFT for the data taken at 175K has several comparable frequency components and does not show any clear frequency. As the measurement temperature is increased, FFT for 200K and 250K data start to show a main frequency component at $6.09 \times 10^{-4}$ Hz, with suppressed background frequencies. FFT analysis for the experimental data taken at 300K, clearly gives the main frequency at $6.09 \times 10^{-4}$ Hz, together with several other frequencies. So as the temperature increased above 175K, the collective oscillations of the nanoparticle system start and after reaching the temperature of 300K, the entire nanoparticle system collectively oscillates with $6.09 \times 10^{-4}$ Hz frequency.

Autocorrelation function, $C(t)$, for the nanoparticles system is calculated next. Fig. 5 shows the calculated values of the $C(t)$ with time, for the experimental data at 175K, 200K, 250K, and 300K. Autocorrelation confirms the observation made through the FFT analysis. Correlation of the system to give oscillations is almost negligible for the data collected at 175K, whereas for the measurements made with increased temperature of 200K and 250K show the enhanced correlated oscillatory behaviour of the system. At 300K, $C(t)$ gives high autocorrelation of the system with the presence of large amplitude oscillations.

Gang et al. (27) and Pikovsky and Kurths (28) have both studied the dynamics of a nonlinear system and established features of stochastic resonance, where the coherent motion of the systems is not stimulated by an external force, but by the intrinsic dynamics of the nonlinear systems. One can see from the autocorrelation plots (Fig. 5) that the correlations are considerably well-defined in the case of the T = 300K data.

We subject the data presented in Fig. 3 to a couple of tests. The first being a standard signal to noise ratio. Temperature as a contributor to stochastic atomic motion has already been employed to demonstrate stochastic resonance (29). In our case, temperature imparts a noise component to the magnetic moment through atomic vibration. We plot the signal as an average of several data records, N, such that $S = 1/N \sum_{i=1,N} \Delta M_i$ with the temperature of observation, T. The results are presented in Fig. 6 where S is seen to rise to a maximum value at 300K, beyond which we have no data as it was technically not possible to operate the SQUID beyond this temperature. This is to an extent correlates with the best degree of coherence resonance (Fig. 3), in the present data, which is observed at 300K. Overall, S is seen to increase with T, whose maximum value is beyond 300K. The second test is based on the theoretical model presented by Pikovsky and Kurths (28) for dynamics of the Fitz Hugh-Nagumo system. If we consider our bistable system as generating magnetic pulses alternately with moment ±M, the separation between the peaks on the positive amplitude side would correspond to activation time $t_a$ for + M and the ones on the negative amplitude side to $t_a$ for - M. While the width of the peak would correspond to $t_e$, the excursion time. A large noise is represented by a decrease in $t_a$. This can be understood in the following manner. According to these authors and their calculation noise enhances or supports the occurrence of pulses, thereby reducing the time of activation between pulses. Hence a decay in $t_a$ is seen as introduction of noise. The $t_a$ for the data presented here is about 3000s (Fig. 3d).We artificially decrease this quantity for data in Fig. 3d, by shifting the entire data by 835s, then adding the resultant data to the original data to obtain a synthetic data. This is shown in Fig. 7(a) and the autocorrelation data derived from the synthetic data, in Fig. 7(b). The synthetic introduction of noise through the activation time parameter has driven the system completely out of coherence resonance.

## Conclusions:

In summary, we have shown thermal noise induced stochastic resonance in the coherent oscillatory behaviour of Fe nanoparticle systems. This coherent resonant phenomenon has been observed due to presence of thermal noise and intrinsic self-organizing magnetic dynamics of the Fe nanoparticle system solely, i.e., without the application of any external periodic force. At 175 K, this system shows oscillations with many frequencies and as the temperature is increased to 300K, the Fe nanoparticle system oscillates collectively with an almost precise frequency. Hence at 300K, the collection of nanoparticles behaves distinctly like a bistable system which generates alternating magnetic pulses with positive and negative magnetic moments. This has been shown to emulate neuron excitations. Above 175K, thermal noise induced ordering of the magnetic moments give rise to the increment of the absolute value of the magnetization and behaves in a manner which can be seen as self-memory storage.

## Acknowledgements

The authors thank Prof. R. Chatterjee of I. I. T., New Delhi for allowing the use of SQUID facility. S.P. Pal thanks CSIR, India, for a research fellowship.

**Figures legends:**

**Figure1.** Schematic representation of: (a) nanoparticles lattice with distorted regions and (b) annealed composite having very few of these regions.

**Figure2.** Magnetic relaxation curves for Fe nanoparticles, taken at: (a) 100K and (b) 175K. Dots and solid lines in curves are experimental data and fitted curves, respectively. Inset in (b), shows periodic oscillations superimposed on the magnetic decay curve.

**Figure3.** Background subtracted oscillatory part for relaxation curves, at: (a) 175K, (b) 200K, (c) 250K, and (d) 300K.

**Figure4.** FFT from magnetic relaxation curves at: (a) 175K, (b) 200K, (c) 250K, and (d) 300K.

**Figure5.** Autocorrelation function for Fe nanoparticles, for the relaxation data taken at: (a) 175K, (b) 200K, (c) 250K, and (b) 300K.

**Figure6.** Calculated value of S for relaxation data taken at different temperatures: 175 K, 200K, 250K, and 300K.

**Figure7.** (a) Synthetic data obtained by 835s shifting, (b) FFT for the data.

**Figures:**

Figure1:

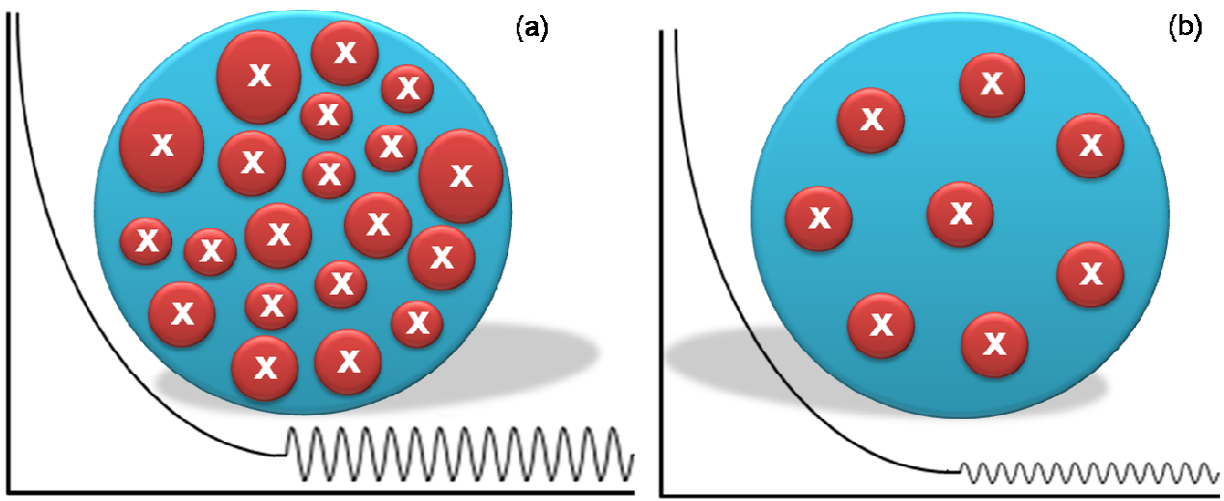

Figure2:

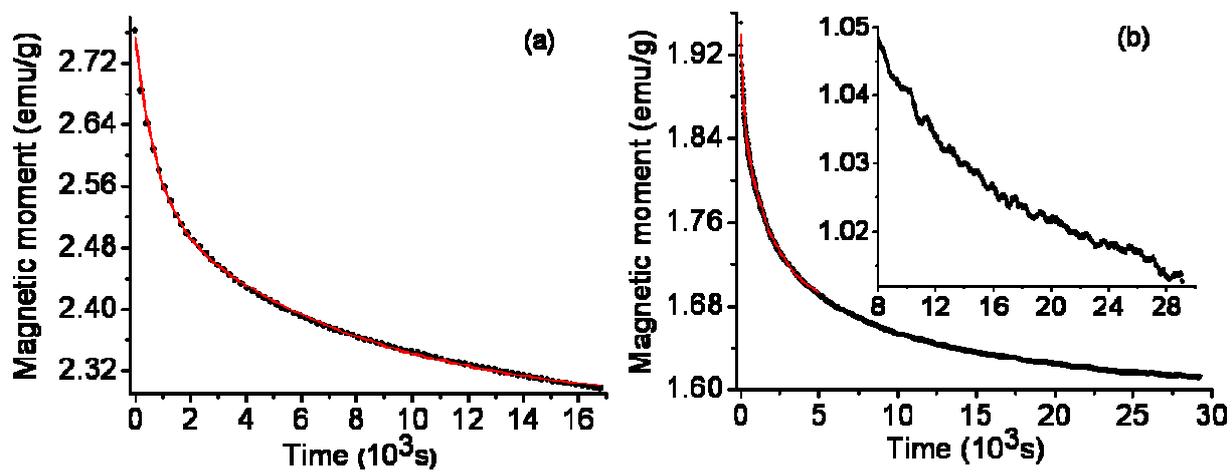

Figure3:

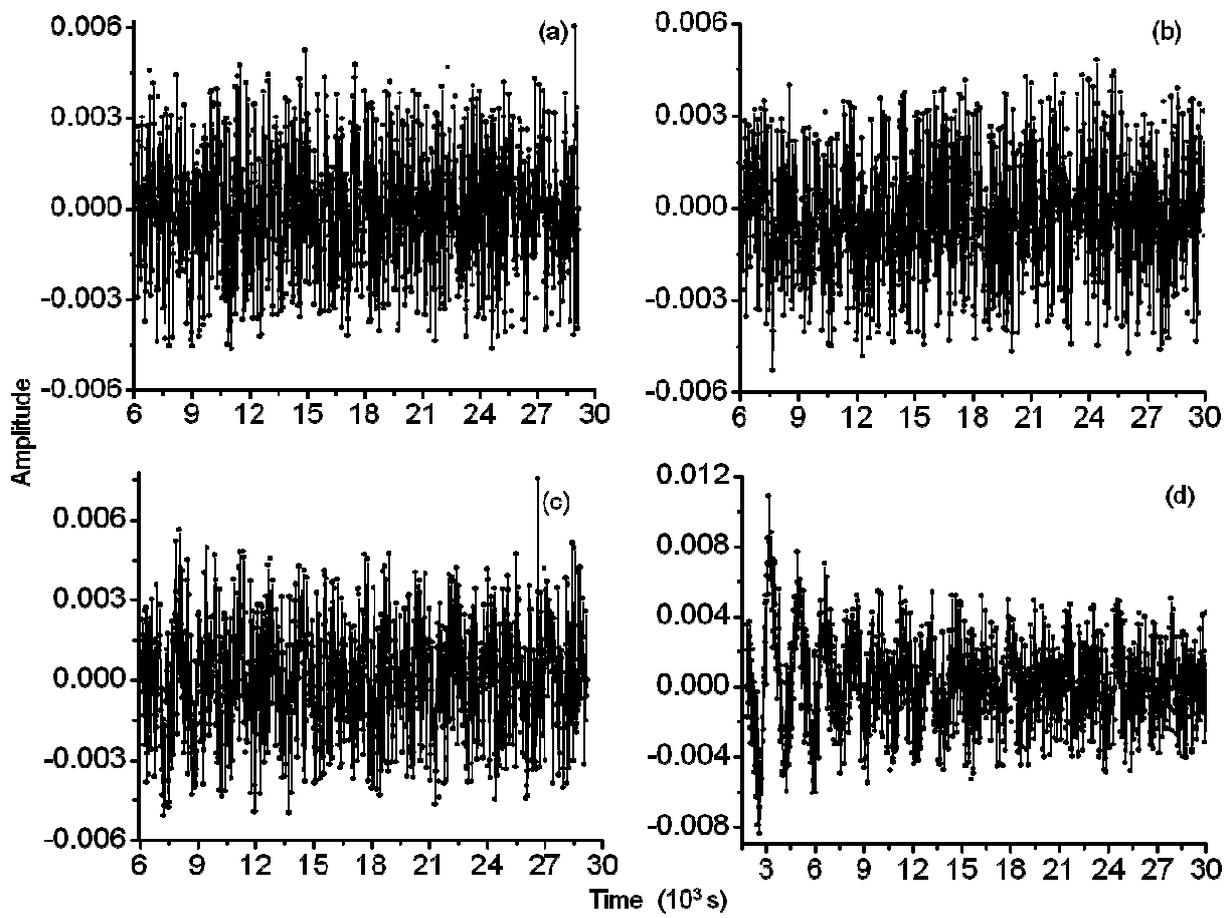

Figure 4:

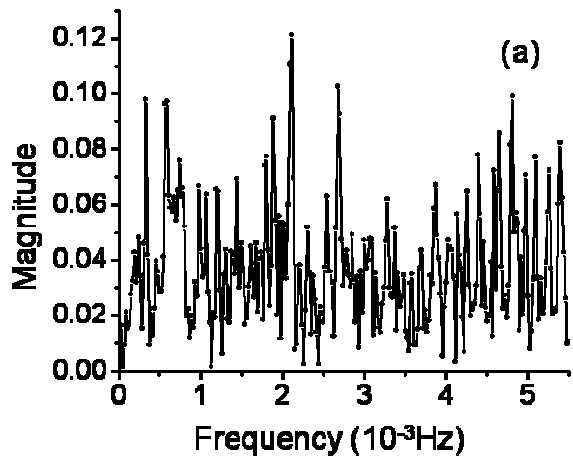
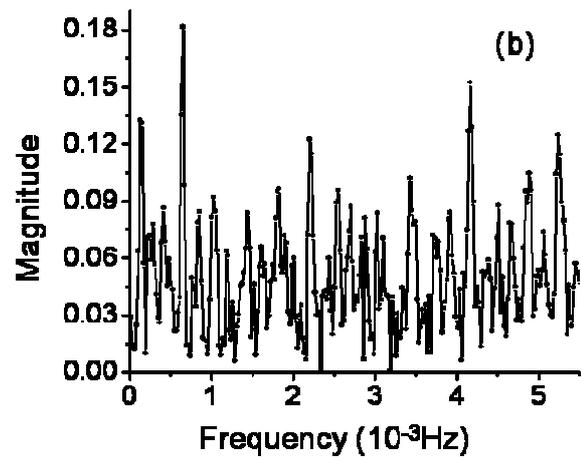
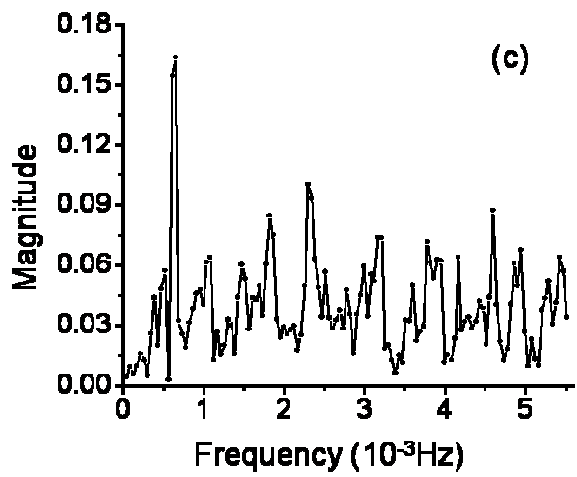
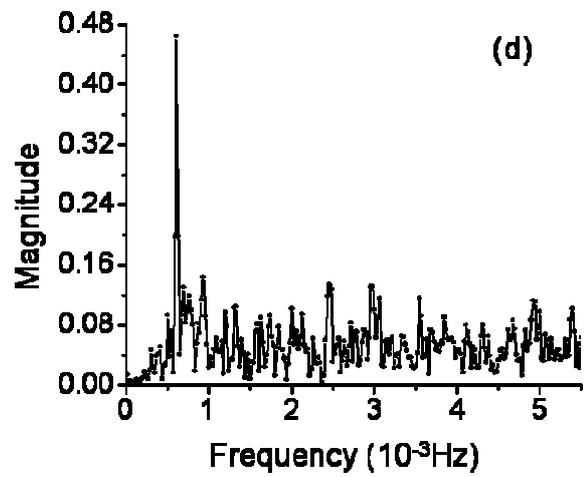

Figure 5:

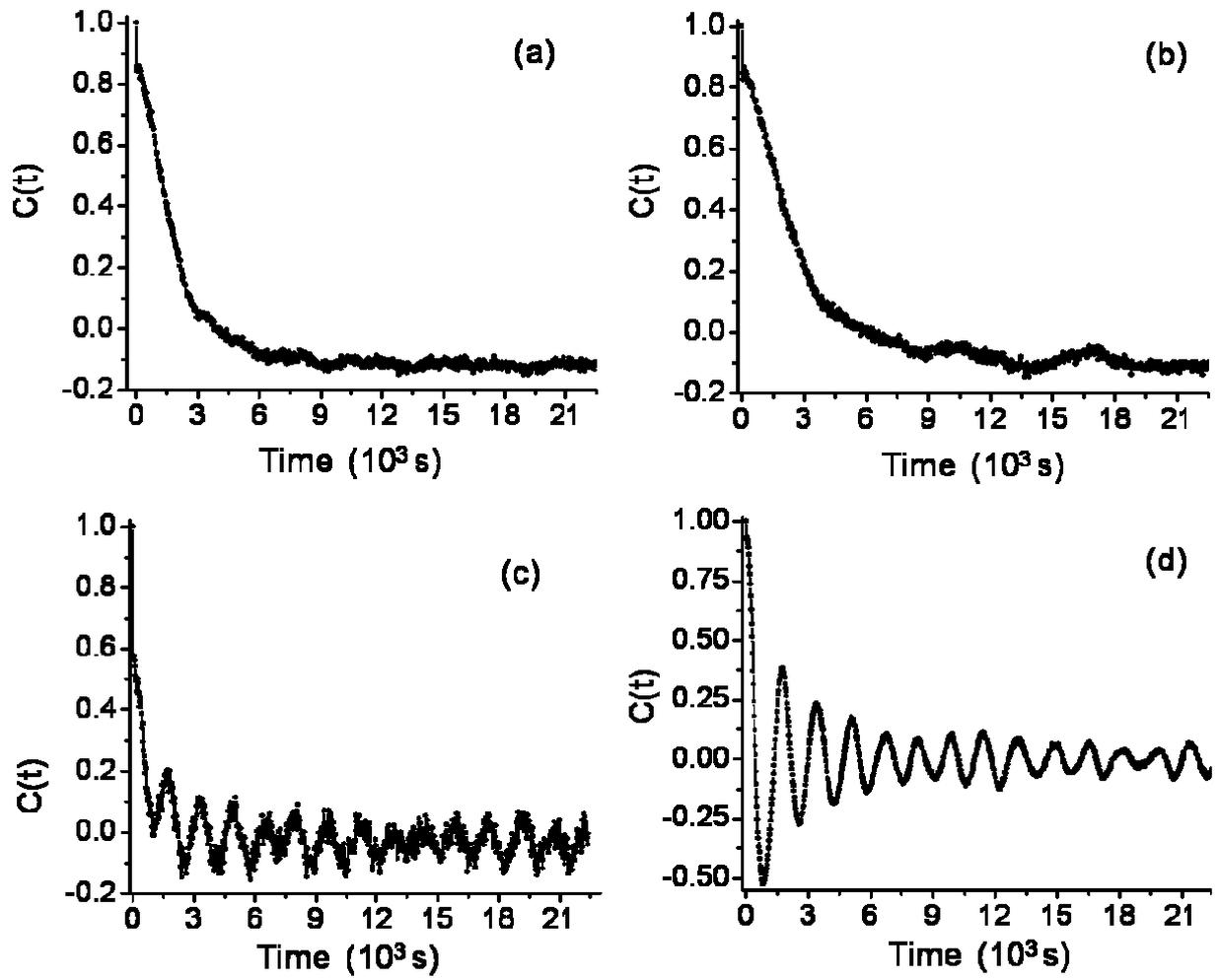

Figure 6:

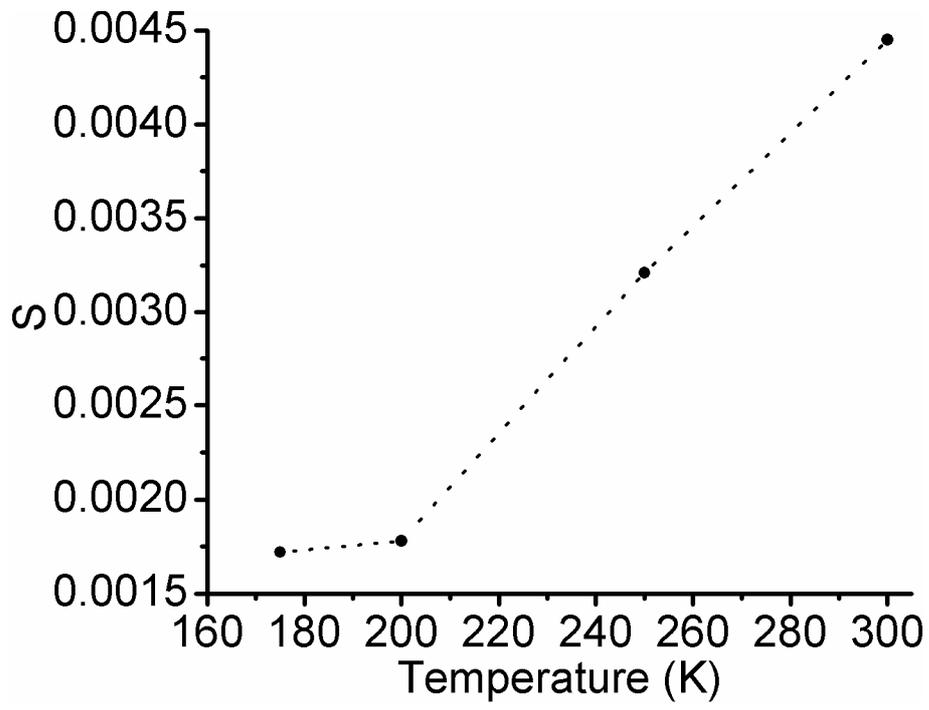

Figure 7:

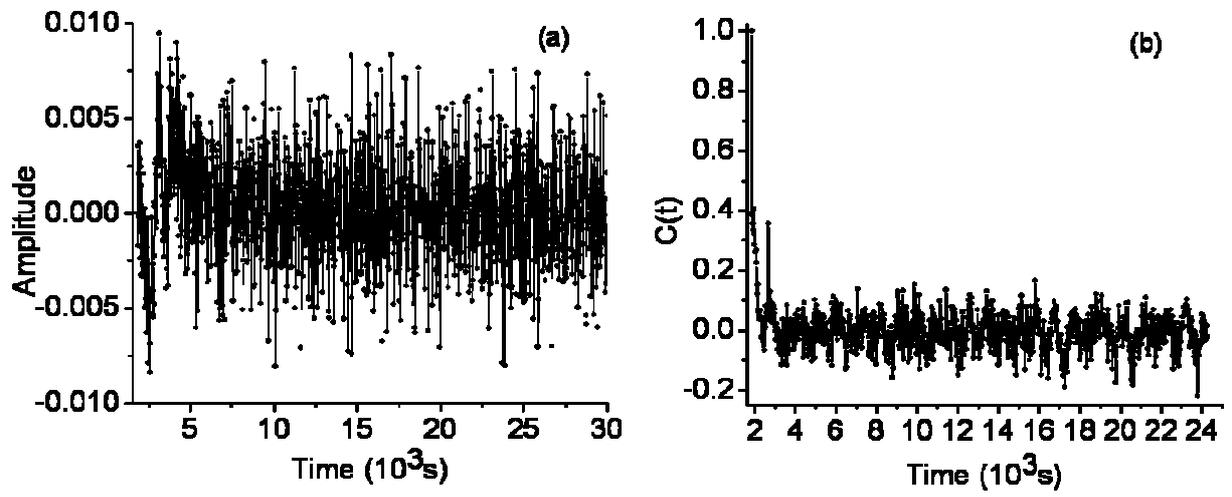